\documentclass[12pt,final]{article} %

\IfFileExists{srcltx.sty}{\usepackage[active]{srcltx}}

\def\hybrid{\topmargin 0pt \oddsidemargin 0pt \headheight 0pt \headsep 0pt
       \textwidth 6.5in        
       \textheight 9in         
           \voffset=-0.3cm         
           \marginparwidth 0.0in \parskip 5pt plus 1pt \jot = 1.5ex}
     
     \hybrid \hfuzz=2pt

\makeatletter
\def\marginnote#1{}

\newcount\hour
\newcount\minute
\newtoks\amorpm
\hour=\time\divide\hour by60
\minute=\time{\multiply\hour by60 \global\advance\minute by-\hour}
\edef\standardtime{{\ifnum\hour<12 \global\amorpm={am}%
           \else\global\amorpm={pm}\advance\hour by-12 \fi
           \ifnum\hour=0 \hour=12 \fi
           \number\hour:\ifnum\minute<10 0\fi\number\minute\the\amorpm}}
\edef\militarytime{\number\hour:\ifnum\minute<10 0\fi\number\minute}

\def\draft{%
  \oddsidemargin -.5truein %
  \def\@oddfoot{\vbox{\scriptsize \hbox{\scriptsize File {\tt\jobname.tex}}
      \hbox{\scriptsize Draft $Revision: 1.11 $}} \hfil { -- \footnotesize
      \thepage\ -- } \hfil \vbox{\scriptsize \hbox to 0pt {\hss\scriptsize
        Time:
            \militarytime}\hbox to 0pt {\hss\scriptsize\today}}%
        \let\@evenfoot\@oddfoot \overfullrule 3pt
}%
}%

\def\numberbysection{\@addtoreset{equation}{section}
  \def\theequation{\thesection.\arabic{equation}}}

\makeatother

\relax 
\hybrid

\numberbysection

\usepackage{amsfonts,varioref,graphicx,xspace} %
\usepackage[intlimits]{amsmath} %
\usepackage{amssymb} %
\usepackage[notcite,notref]{showkeys}

\usepackage[colorlinks,bookmarks=false]{hyperref}

\renewcommand{\vref}[1]{\ref{#1}~\vpageref{#1}\unskip\xspace}
\newcommand{\p}[1]{\partial_{#1}} %
\newcommand{\la}{\langle} %
\newcommand{\ra}{\rangle} %
\newcommand{\ket}[1]{\left|{#1}\right\rangle} %
\newcommand{\bra}[1]{\left\langle{#1}\right|} %
\newcommand{\zm}{{\mathrm{z.m}}} 
\newcommand{\cs}{{\textsc{cs}}} 
\newcommand{\z}{{\textsc{z}}} 
\newcommand{\anom}{{\scriptstyle{\textsc{anom}}}} %
\newcommand{\Y}{{\scriptstyle{\textsc{y}}}} %
\newcommand{\Tr}{\mathop\mathrm{Tr}}  
\newcommand{\sign}{\mathop\mathrm{sign}} 

\newcommand{\e}{\,\mathbf{e}} 
\newcommand{\bE}{\text{\textbf{\slshape E}}} 
\newcommand{\bH}{\text{\textbf{\slshape H}}} 
\newcommand{\M}{M} 
\newcommand{\m}{\text{\textbf{\slshape m}}} 
\newcommand{\hs}{\hskip -.25em}
\def\CF{{\cal F}} %
\def\CL{{\cal L}} %
\begin{document}
\title{Anomalies as a signature of extra dimensions} %

\author{\large \sf Alexey Boyarsky\thanks{Ecole Polytechnique F\'ed\'erale de
    Lausanne, Institute of Theoretical Physics, FSB/ITP/LPPC, BSP 720,
    CH-1015, Lausanne, Switzerland},~~Oleg Ruchayskiy\thanks{Institut des
    Hautes \'Etudes
    Scientifiques, Bures-sur-Yvette, F-91440, France},~~Mikhail Shaposhnikov$^*$.} %

\date{} \maketitle
\begin{abstract}
  The anomaly cancellation condition of the Standard Model may be unnatural in
  theories with extra dimensions as an anomaly of a low-energy 4-dimensional
  theory can be canceled by an inflow from a bulk.  This inflow may give rise
  to an observable effect at low energies.  We analyze several physical models
  in which this effect exists and estimate constraints on its value, imposed
  by the modern experimental data.  We show that the effect can be large
  enough to be observed even when these constraints are satisfied.  Positive
  result of such an experiment would be a low-energy signature of the
  existence of extra dimensions.
\end{abstract}

\section{Introduction} 
\label{sec:intro} 

Models with extra dimensions are interesting candidates to describe physics
beyond the Standard Model (SM). It is generally assumed that such a new
physics appears at high energies, say at 1 TeV or above. It is natural to ask
whether there could exist some low energy signatures of extra dimensions.  In
this paper we will show, that in addition to a possible modification of
gravitation Newton's law at distances of a fraction of
millimeter~\cite{mm-gravity}, there could be quantum effects, related to the
anomalies of chiral gauge theories.  Anomalies do not depend on a scale (e.g.
the scale of the extra dimensions) which makes possible that corresponding
effects are visible at low energy.

In the Standard Model in 4 dimensions the fermion couplings are chiral, and
the hypercharges of quarks and leptons are chosen in such a way as to ensure
cancellation of all gauge anomalies~\cite{sm-anom}. However, in theories with
extra dimensions, where the SM fields are realized as zero modes of a
higher-dimensional ones, only the full (higher-dimensional) theory should be
anomaly free and there is no reason to expect separate anomaly cancellation
for the brane fields (for an explicit example of a consistent brane-world
theory with an anomaly on the brane, see~\cite{Shap.qed}). The aim of this
paper is to study exactly this type of models.

Qualitatively, if the theory on the brane is anomalous, then there is a
specific type of \emph{brane-bulk interaction} -- \emph{anomaly
  inflow}~\cite{ch}. It is described by Chern-Simons-like terms in the
low-energy effective action in the bulk.  These terms are not gauge invariant
in the presence of a brane. Therefore, they generate currents, flowing to the
brane and ensuring the gauge invariance of the full theory. Such a mechanism
appears in many problems: in field theories with solitonic backgrounds (see
e.g.~\cite{hr,Shap.qed}), in string and M-theory~\cite{fhmm}.

What would be the consequences of the presence of inflow currents
from the point of view of a low energy physics on a brane?  For
example, can one distinguish between a 4-dimensional anomaly-free
theory and a theory with extra dimensions and anomaly inflow? 

Naively, one would think that anomaly inflow current is a flux of particles
from higher dimensions. From 4-dimensional point of view such a process would
seem as a ``loss of unitarity''. However, for a brane-world model to be
phenomenologically acceptable, there should be a mass gap between zero and
bulk modes of matter fields and zero mode particles cannot leave the brane at
energies below the mass gap.  Instead, the inflow current by its nature is a
\emph{vacuum} or \emph{non-dissipative} current. Although inflow current is
not carried by real particles, the redistribution of the charges in the Dirac
sea of the full (5-dimensional) theory leads to an appearance of an electric
charge on the brane~\cite{anomaly-th}. As a result, for example, electric
field, created by a capacitor, changes when the latter is placed in a magnetic
field. In particular, a field appears outside the capacitor, while inside it
diminishes in value and gets redistributed in space.  Another example is a
charged elementary particle: when placed in a magnetic field its electric
charge gets screened. From the 4-dimensional point of view it looks as if
\emph{photon had acquired a mass, depending on a magnetic field}. In addition,
particle with both electric charge and magnetic moment acquires time-dependent
dipole moment.  These effects would not be present in a purely 4-dimensional
theory with anomaly canceled by addition of new (chiral) particles at higher
energies, and thus can be considered a signature of extra dimension. As we
will see, they are \emph{not} suppressed by the mass scale of extra
dimensions.

In the present paper we study the physical manifestations of the
above effects and suggest an experiment in which they can be
observed.\footnote{We analyze only gauge anomalies here. Brane-world
models with global chiral anomaly has been studied
in~\cite{theta-vacuum}. They may lead to the solution of the strong
CP problem.} The paper is organized as follows. We describe the
observable effects of anomaly inflow in Section~\ref{sec:qed} and
then turn in Section~\ref{sec:anom-sm} to the question of possible
"anomalous" extensions of SM. Next, we discuss an experiment in which
the screening of electric charge can be observed. In
Section~\ref{sec:numbers} we analyze several physical models and
obtain quantitative estimates of the screening.

\section{Anomalous electrodynamics and its observational\\ signatures.}
\label{sec:qed}

\subsection{5-dimensional electrodynamics on a brane}
\label{sec:qed5d}

To demonstrate desired effects we start with a $U(1)$ theory in 5 dimensions.
This model was discussed in details in~\cite{anomaly-th}, here we only remind
its main features. We consider a 5-dimensional brane-world which has both
chiral fermions and gauge fields localized on a brane.  The low energy
effective action of this theory consists of three parts: kinetic term for the
gauge fields; Chern-Simons like term, describing a brane-bulk interaction,
leading to inflow; and an action of (chiral) fermionic zero modes, localized
on the brane (hence $\delta(z)$ in the last
term):\footnote{
  Our conventions are as follows: Latin indices $a,\dots,e=0,\dots,4$, Greek
  $\mu,\nu= 0,\dots,3$. We are using \emph{mostly negative} metric. Our brane
  is stretched in $0,\dots,3$ directions and is located at $z = 0$ (where
  $z\equiv x^4$). We will use notations $t,x$ for $x^0,x^1$ and choose polar
  coordinates $(r,\theta)$ in the plane $(x^2,x^3)$. Spatial coordinates on
  the brane $(x^1,x^2,x^3)$ will somtimes be denoted by $\vec
  x$.\label{fn:1}}%
$^,$\footnote{We choose all the characteristic energy scales in the problem to
  be much less than the mass gap of the fermionic zero modes $m_\psi$,
  therefore their profile can be approximated by the delta-function
  $\delta(z)\approx 2m_\psi e^{-m_\psi|z|}$, which appears in
  eqs.~(\ref{eq:4}),~(\ref{eq:5}), etc.}
\begin{equation}
  \label{eq:4}
  S = -\frac1{4\e^2}\int d^5 x\, \Delta(z) F_{ab}^2 + \frac14\int
  d^5x\,\kappa(z)\epsilon^{abcde}A_a F_{bc} F_{de} + \int\hs d^5x\,
  \CL_\zm\delta(z) \:.
\end{equation}
Here $\e$ is a five-dimensional charge, with the dimensionality of
$(\mathrm{length})^{\frac12}$.  Localization on chiral fermions is achieved as
in e.g.~\cite{rub-shap}.  To localize $U(1)$ gauge field we modified kinetic
term in~(\ref{eq:4}): a $U(1)$ gauge field will possess a massless zero mode
on the brane, if the factor $\Delta(z)$ in the kinetic term is chosen properly
(we refer for details
to~\cite{oda,dubovsky,st,gauge-fields,anomaly-th,dvali-localization}).

The charges in the full theory are chosen such that the theory of fermion zero
modes (described by $\CL_\zm$) suffers from a gauge anomaly. Namely, the
electric current of these modes $j^\mu_\zm \equiv \frac{\delta \CL_\zm}{\delta
  A_\mu}$ is not conserved:
\begin{equation}
  \label{eq:5}
  \p\mu j^\mu_\zm = \frac32\kappa_0
  \epsilon^{\mu\nu\lambda\rho}F_{\mu\nu}F_{\lambda\rho}\,\delta(z) \:.
\end{equation}
Parameter $\kappa_0$ is proportional to the difference of the (cubes) of the
charges of left and right-handed fermions~\cite{abj,anomaly-th}. From a
4-dimensional point of view such a theory would be inconsistent at quantum
level~\cite{gross-jackiw}.  However, being embedded in a 5-dimensional
theory~(\ref{eq:4}), it becomes perfectly well-defined. This is due to the
special type of an interaction (\emph{anomaly inflow}~\cite{ch}) represented
by a Chern-Simons-like term in~(\ref{eq:4}).
Compared to the usual cases when Chern-Simons terms appeared in an effective
action~\cite{redlich}, this term has an additional factor $\kappa(z)$.  In the
limit of an infinitely thin wall and the large fermionic mass gap it reduces
to $\kappa(z) = \kappa_0\sign(z)$. Due to the presence of the brane, which
plays the role of a boundary, Chern-Simons term is not gauge invariant and its
variation cancels that of $\CL_\zm$. Namely, the divergence of the
Chern-Simons current is opposite to that of $j_\zm^\mu$:
\begin{equation}
  \label{eq:9}
    \p\mu j_\zm^\mu = - \p a J_\cs^a \neq 0\:,
\end{equation}
where the inflow current $J_\cs$ is defined via\footnote{The term
  in~(\ref{eq:8}), proportional to the $\kappa'(z)$, can be thought of as a
  local counterterm, making the current of zero mode $j_\zm^\mu$ anomalous.
  See~\cite{cov-anom,hr}.} %
\begin{equation}
  \label{eq:8}
  J^a_\cs(x^a) = \frac{\delta S_\cs}{\delta A_a} = 
  \frac{3}{4}\kappa(z)
  \epsilon^{abcde}F_{bc}(x^a)F_{de}(x^a)+\frac12\kappa'(z)\delta^{a}_\mu 
\epsilon^{\mu\nu\lambda\rho} A_\nu F_{\lambda\rho},\quad
  \kappa'(z) =\frac{d \kappa(z)}{d z}\:.
\end{equation}

\noindent If the theory were purely four-dimensional, non-zero divergence~(\ref{eq:5})
of the electric current $j^\mu_\zm$ would have described the rate of anomalous
particle production, essentially the measure of non-unitarity of the theory.
In the 5-dimensional brane-world this divergence gets canceled by that of the
Chern-Simons (or \emph{inflow}) current $J^a_\cs$ (eq.~(\ref{eq:9})).
According to the definition~(\ref{eq:8}) for configurations of 4-dimensional
electro-magnetic fields, such that
$\epsilon^{\mu\nu\lambda\rho}F_{\mu\nu}F_{\lambda\rho}\neq0$, there exists the
component $J^z_\cs$, i.e the Chern-Simons current ``flows onto the brane''.
However, it would be wrong to think about this current as ``bringing to the
brane particles from extra dimension''. In reality this is a \emph{vacuum
  current} -- a collective effect of redistribution of the bulk (massive)
fermions.  At the energies of interest these fermions cannot be excited from
the vacuum and cannot create actual charge carriers. The effect of the
redistribution of the particles in the vacuum, however, does not depend on
their mass and thus inflow interaction is not suppressed by the mass gap.
(This fact can be checked by the direct microscopic computations of the vacuum
average of $\bra 0 J_\cs^z\ket 0$ in the full theory).  One may think of the
effect of the vacuum current as of a dielectric susceptibility of the vacuum
of four-dimensional theory, embedded in the 5-dimensional space-time.

We see that the main characteristic feature of theories with anomaly on the
brane, is the presence of the vacuum inflow current from extra dimensions. The
presence of such a current exhibits itself at low energies, as we discuss
below.

\subsection{Anomalous electric field of a capacitor}
\label{sec:observ-inflow}


To design a setup in which anomaly inflow could exhibit itself we need a
configuration of (five-dimensional) gauge fields in which
$\epsilon^{\mu\nu\lambda\rho}F_{\mu\nu}(x^a)F_{\lambda\rho}(x^a)|_{z=0}\sim
\vec E\cdot \vec H$ is non-zero (we will take $F^{01}= E$ and $F^{23}= H$ in
suitably chosen coordinate system).  We realize such fields by two parallel
infinite plates in the vacuum, separated in the direction $x$ by the distance
$2d$ -- a capacitor with initial charge densities $\pm\sigma_0$ on plates,
placed in the magnetic field. Magnetic field is created by an (infinite in the
$x$-direction) solenoid with the radius $R$. Below we report the results, for
details a reader is referred to~\cite{anomaly-th}.

A final stage of anomaly inflow in this system is such that the static
solution (non-perturbative in $\kappa_0$) will form.  The main property of
this solution is that an \emph{electric field gets screened as if photon had
  become massive} with the mass
\begin{equation}
  \label{eq:49}
  m_0^2 \sim \kappa_0 \bH_0 (\!\e^2  \M) =\kappa_0\, e^2 \bH_0 
\end{equation}
(where $e^2$ is a \emph{4-dimensional coupling constant}, related to the
5-dimensional $\e^2$ via $e^2 \sim \e^2 \M$). The mass $m_0$ depends only on
the 4-dimensional quantities (fine-structure constant $e^2$ and the magnetic
field $\bH_0$ as seen by an observer on a brane).  Notice, that $m_0\ll \M$
for any physically plausible values of magnetic field $\bH_0$.\footnote{Recall
  that 1~Gauss$=1.95\cdot 10^{-20}$~GeV$^2$ while $\M\gtrsim 10^4$~GeV. For
  the maximal values of magnetic fields, see Section~\ref{sec:numbers}.}

The profile of a 4-dimensional electric field in the static state is given by
\begin{equation}
  \label{eq:54}
  \boxed{
    \bE^x(x) = -\frac{\bE_0}2  \Bigl[\sign (x-d)e^{-m_0 |x-d|} -
      \sign(x+d) e^{-m_0 |x+d|}\Bigr]}\:.
\end{equation}
Here $\bE_0$ is a value of 4-dimensional field, which would be created between
the plates of the capacitor in the theory without anomaly $\bE_0 = \sigma_0
e^2$.  The characteristic time $\tau_0$ over which the solution~(\ref{eq:54})
is established is of the order $m_0^{-1}$.  Of course, as
$m_0\to0$~(\ref{eq:54}) reproduces an electric field of an infinite capacitor.
The full (5-dimensional) electrostatic potential, corresponding to the
solution~(\ref{eq:54}) has the following form:
\begin{equation}
\label{eq:51}
  \Phi(x,z) = \phi_0(x) \chi_0(z)=-\frac{\e^2 {\sigma_0}}{2m_0} \left(
    e^{-m_0 |x-d|} -    e^{-m_0 |x+d|}\right) \chi_0(z)\:,
\end{equation}
with the $x$ profile $\phi_0(x)$ satisfying Poisson equation with mass $m_0$:
$ \p x^2 \phi_0(x) - m_0^2 \phi_0(x)=
\sigma_0\Bigl(\delta(x-d)+\delta(x+d)\Bigr)$.
 
The profile of the solution~(\ref{eq:51}) in the $z$ direction $\chi_0(z)$ is
sharply localized
in the region\footnote{
  In case of the warp factor $\Delta(z)=\exp(-2\M|z|)$~\cite{anomaly-th} the
  profile function $\chi_0(z)$ is given by $\frac{m_0 e^{\M|z|}}{\sqrt{6\pi}}
  K_{\nu}\left( \tfrac{m_0^2}{4\M^2} e^{2\M|z|}\right)$ where $K_\nu$ is the
  modified Bessel function of the second kind with the index $\nu =
  \frac12\sqrt{1-{m_0^2}/{\M^2}}$ (we assume that $m_0 < \M$).  Therefore
  outside the region~(\ref{eq:53}) potential $\Phi$ decays as an exponent
  $e^{-\Delta(z)}$.  Notice, that $K_\nu$ is non-perturbative as a function of
  $\kappa_0$.}
\begin{equation}
  \label{eq:53}
  |z| \lesssim 
  \frac1{\M}\log\frac {\M}{m_0} \approx\frac1{2\M}\log\frac{\M^2}{e^2 \kappa_0 \bH_0}\:.
\end{equation}
Thus, the function $\Phi(x,z)$ exponentially decays in $z$ direction on the
scale proportional to the $\M^{-1}$ and depending on $\kappa_0$ (potentially
very small) only logarithmically, while the scale $m_0^{-1}$ of the
exponential decay of the potential in the $x$ direction \emph{does not depend
  on $\M$} and is proportional to the $\sqrt{\kappa_0}$. 

\textbf{To summarize:} an effect of anomaly inflow exhibits itself via a
change of a distribution of an electric field in a capacitor.  The field
diminishes inside
\begin{equation}
  \label{eq:55}
  \bE^x_{inside} = \bE_0 e^{-m_0 d}\cosh m_0 x,\quad |x| < d\:,
\end{equation}
but appears outside:
\begin{equation}
  \label{eq:56}
  \bE^x_{outside} = -\bE_0 e^{-m_0 (|x| -d)}(1 - e^{-2 m d}), \quad |x|> d \:.
\end{equation}
If $\kappa_0$ is such that $m_0 d\ll 1$, the field outside the capacitor (for
$ \frac1{\M}\ll(|x|-d)\ll \frac1{m_0}$) is almost constant, given by
\begin{equation}
  \label{eq:57}
  \bE^x_{outside} \approx 2\bE_0 (m_0 d)\:.
\end{equation}


\section{The Standard Model with anomaly inflow}
\label{sec:anom-sm}

In this section we apply the logic of Section~\ref{sec:qed} to SM.  Indeed, if
the SM fields are localized on a brane in a 5-dimensional world, there is no
apparent reason to expect a separate anomaly cancellation for them.  For
simplicity we consider the electroweak $SU(2)\times U(1)$ theory with only one
generation of fermions (an addition of extra generations does not change the
analysis). The action for the $SU(2)$ and $U(1)$ gauge fields is similar to
that in eq.~(\ref{eq:4}). The warp-factor $\Delta(z)$ is such that there is a
zero mode, localized on the brane. We also add a Higgs $\phi$ field, which is
an $SU(2)$ doublet. Its mass $m_\phi^2(z)$ is negative at $z=0$ and tends to
the positive constant in the bulk as $|z|\to \infty$. Therefore its VEV has a
non-trivial profile $\la\phi(z)\ra$, with $\la\phi(0)\ra\neq 0$ and
$\la\phi(z)\ra\to 0$ at $z\to\pm \infty$.  So the $SU(2)\times U(1)$ symmetry
is broken down to $U(1)_{\textsc{em}}$ on the brane and restored far from it.
As a result, the vector boson acquires mass $\m_\z$ on the brane, while
remains massless at large $|z|$.

We add fermions in the bulk, charged with respect to $SU(2)\times U(1)$.
Fermionic zero modes are localized on the brane. 
There is a Yukawa interaction of these modes with the Higgs field $\phi$:
\begin{equation}
  \label{eq:88}
  {\cal L}_{Yukawa} = \lambda_e \bar E_L^a \phi^a e_R  + \lambda_d \bar
  Q_L^a \phi^a d_R + \lambda_u \epsilon_{ab}\bar Q_L^a  (\phi^+)^b u_R  +
  \lambda_\nu \bar E_L^a \epsilon_{ab}(\phi^+)^b \nu_R\:,
\end{equation}
which leads to mass terms of brane fermions after spontaneous symmetry
breaking.\footnote{We use here the standard notations: left-handed fermions
  ($\nu_L$, $e_L$, $u_L$, $d_L$) form a doublet with respect to the $SU(2)$
  group, right-handed particles ($e_R$,
  $u_R$, $d_R$) are singlets.} %
We can also add a right-handed neutrino $\nu_R$ with a Majorana mass term.
The charges of these particles with respect to the $U(1)_\Y$ group
(\emph{hypercharges}) are related due to the structure of the
interaction~(\ref{eq:88}):
\begin{equation}
  \label{eq:89}
  \left\{
    \begin{array}{rcl}
      Y_{E} & = &Y_\phi + Y_{e}  \:,\\
      Y_{Q} & = &Y_\phi + Y_{d}  \:,\\
      Y_{Q} & = &Y_{u}- Y_\phi  \:.
    \end{array} %
  \right.
\end{equation}
Hypercharges and \emph{isospins} (charges with respect to the $SU(2)$) of
particles are related to their electric charges via
\begin{equation}
  \label{eq:90}
  Q_{\textsc{em}} = \tfrac12\sigma^3 + \tfrac Y{ 2 Y_\phi}
\end{equation}
(here $\sigma^3$ is a Pauli matrix).  We will choose the normalization $Y_\phi
=1$ from now on.

Notice, that for any choice of hypercharges, consistent with~(\ref{eq:89}),
electrodynamics remains vector-like. One could (erroneously) conclude that we
cannot expect to detect any consequences of anomaly inflow at low energies, as
only electromagnetic background fields can be created, while the corresponding
current is non-anomalous. However, we will show in the following sections that
this is not the case, and in fact we can observe effect, identical to that of
the Section~\ref{sec:observ-inflow}.

\subsection{Charge difference of electron and proton and anomalies}
\label{sec:anom-charges}

\begin{table}[t]
  \begin{center}
    \begin{tabular}{|c|c|c|c|c|c|c|}
      \hline
      Particle: & $e_L$ & $e_R$ & $\nu_L$ & $Q$ & $u$ & $d$\\
      \hline
      Hypercharge $Y$:& $-1+\kappa_l$  & $-2+\kappa_l$& $-1+\kappa_l$ &
      $\frac13+\kappa_q 
      $& $\frac43+\kappa_q$& $-\frac23+\kappa_q$\\ 
      \hline
    \end{tabular} $+$
    \begin{tabular}{|c|}
      \hline
      $\nu_R$\\
      \hline
      $\kappa_l$\\
      \hline
    \end{tabular}
    \caption{Modification of hypercharge assignments in SM. Standard
      assignments are shifted by $\kappa_l$ in lepton sector and by $\kappa_q$
      in quark sector. The right-handed neutrino can be added to cancel also
      gravitational anomaly, but does not affect our analysis otherwise.}
    \label{tab:1}
  \end{center}
\end{table}
We consider the extension of SM in which one chooses
\emph{arbitrary values of hypercharges}, consistent with~(\ref{eq:89}).  This
extension can be described in terms of two parameters $\kappa_l$ and
$\kappa_q$, which shift the standard values of hypercharges in lepton and
quark sectors correspondingly (see Table~\ref{tab:1}). We are going to show
that all anomalies, which arise as a result of our choice of hypercharges,
will be proportional to the same combination of $\kappa_l$ and $\kappa_q$,
which is equal to the charge difference between the electron and the proton.

For generic values of $\kappa_l$ and $\kappa_q$ in Table~\ref{tab:1} the
$U(1)_Y$ hyper-current $j_\Y^\mu$ becomes anomalous\footnote{The gravitational
  anomaly (triangular diagram with one hypercurrent and two gravitons) is
  proportional to the trace over all hypercharges $\Tr[Y]$ and does not arise
  if one adds a right-handed neutrino: $\Tr[Y]=0$.}
\begin{equation}
  \label{eq:92}
  \p\mu j_\Y^\mu =  \frac{\Tr[Y^3]}{16\pi^2}
  \epsilon^{\mu\nu\lambda\rho}\CF_{\mu\nu}\CF_{\lambda\rho}
  +\frac{\Tr[Y_L]}{16\pi^2}   \epsilon^{\mu\nu\lambda\rho}\Tr
 G_{\mu\nu} G_{\lambda\rho}\:.
\end{equation}
Here $\CF_{\mu\nu}$ is a $U(1)_\Y$ field strength of the $U(1)$ field $B_\mu$
(hyper-photon): $\CF_{\mu\nu} = \p\mu B_\nu - \p\nu B_\mu$;
$G^\alpha_{\mu\nu}$ is an $SU(2)$ non-Abelian field strength, $g'$ and $g$ are
$U(1)_\Y$ and $SU(2)$ coupling constants correspondingly.  The first term
in~\eqref{eq:92} comes from the triangular diagram with three $U(1)$ vertices
and is proportional to the sum of cubes of hypercharges of \emph{all}
particles.  The second term comes from the diagram with one $U(1)_\Y$ and two
$SU(2)$ vertices.  It is proportional to the sum of hypercharges of all left
particles -- those, having non-trivial $SU(2)$
charges.  Traces in~(\ref{eq:92}) are given by
\begin{equation}
  \label{eq:93}
  \Tr[Y^3] = 6(\kappa_l+3\kappa_q)\qquad \text{and}\qquad 
  \Tr[Y_L] =-2 (\kappa_l+3\kappa_q)\:. 
\end{equation}
Along with anomaly~\eqref{eq:92} there is also a non-conservation of $SU(2)$
current in the background $U(1)$ and $SU(2)$ fields (its coefficient is
proportional to the same diagram as the second term in~\eqref{eq:92}):
\begin{equation}
  \label{eq:94}
    D^\mu j_\mu^\alpha =  \frac{\Tr[Y_L]}{8\pi^2}
  \epsilon^{\mu\nu\lambda\rho}G^\alpha_{\mu\nu} F_{\lambda\rho}\:.
\end{equation}
We see, that in our extension of the electroweak theory all anomalies are
proportional to the same combination of $\kappa_l$ and $\kappa_q$:
\begin{equation}
  \label{eq:95}
  \kappa \equiv \kappa_l+3\kappa_q\:,
\end{equation}
which we will call \emph{anomaly coefficient} in what follows. In the absence
of right-handed neutrino, the requirement of cancellation of both
anomalies~(\ref{eq:92}) and~(\ref{eq:94}) would lead to the standard
hypercharge assignments of SM (namely, $\kappa_l =
\kappa_q = 0$).  With the right-handed neutrino being present, requirement
of anomaly cancellation would only fix $\kappa=0$, and thus impose
$\kappa_l = -3\kappa_q$ condition.\footnote{Experimental restrictions for
  $\kappa_l$ come from the measurements of the electric charge of neutrino:
  $Q_\nu /Q_e= \kappa_l \sim 10^{-15}$, while the estimates for $\kappa_q$
  are due to the measurements of the charge of the neutron: $Q_n/Q_e =
  3\kappa_q\sim
  10^{-20} $~\cite{pdg}.}
$^,$\footnote{One should check whether the electric charge of neutrino can
  lead to observable effects, e.g.  pair production on neutrinos in the strong
  electric field.  Probability of such production is given by: $ P_\nu \sim
  \exp\left(-\frac{\pi m_\nu^2}{Q_\nu E}\right) $. For the maximum electric
  field $E\sim 10^7\, \mathrm{Volt}/\mathrm{cm}$, taking electric charge to be
  $Q_\nu = 10^{-15}Q_e$ and mass $m_\nu \sim 1\,\textrm{eV}$, we get $P\sim
  e^{-10^7}$.  The effect of neutrino production becomes visible only if mass
  of neutrino is $m_\nu \lesssim 10^{-3}\,\mathrm{eV}$.} 

The expression~\eqref{eq:95} is equal to the difference of (the absolute
values of) electric charges of electron and proton and is experimentally
measurable quantity.  Current experimental bound puts $\kappa<10^{-21}$
~\cite{charge-diff}.
If SM is embedded into a higher-dimensional anomaly-free
theory, \emph{a priori} one has no reason to expect $\kappa=0$ and the fact
that its value is so small experimentally becomes a fine-tuning, which a realistic
brane-world model should be able to explain.

Recall that $U(1)_Y$ and $SU(2)$ fields, entering~\eqref{eq:92}
and~\eqref{eq:94}, are the fields \emph{above} the electroweak symmetry
breaking scale.  At lower energies it is convenient to re-express these
anomalies in terms of of electro-magnetic field $\gamma_\mu$ and
\emph{neutral} field $Z_\mu$, which can be obtained from the $U(1)$ and
$SU(2)$ fields after rotation by the Weinberg angle $\theta_w$.\footnote{One
  should take into account that we are using a normalization of the gauge
  fields with a coupling constant in front of the kinetic term.}
%
The electro-magnetic current $j_{\gamma}^\mu=j^\mu_\Y+j_3^\mu$ and
\emph{neutral current} $j_{\z}^\mu = -\cot\theta_w j_\Y^\mu+\tan\theta_w
j^{\mu}_3$, where $j^{\mu}_3$ is the 3rd component of the $SU(2)$ triplet
$j_\alpha^\mu$, $\alpha=1,2,3$.  Using~(\ref{eq:92}) and~(\ref{eq:94}) one can
easily see that \emph{(i)} electro-magnetic current is conserved in the
arbitrary background of electro-magnetic fields (as one expected, because the
electrodynamics remains vector-like in our model); \emph{(ii)} there is an
anomalous $\gamma\gamma Z$ \emph{coupling}
\begin{equation}
  \label{eq:96}
  \p\mu j^\mu_\z = -\frac{2\,\,N_f\,\kappa}{\pi^2\cos\theta_w\sin\theta_w}
  \vec E_\gamma\cdot  \vec H_\gamma\:,
\end{equation}
which implies the non-conservation of the neutral current in the parallel
electric and magnetic fields ($N_f$ is the number of generations);
\emph{(iii)} another important consequence of the presence of $\gamma\gamma
\z$ coupling is the non-conservation of the electro-magnetic current in the
mixed electro-magnetic and $\z$-backgrounds:
\begin{equation}
  \label{eq:97}
  \p\mu j_\gamma^\mu =  -\frac{4\,N_f\,\kappa}{\pi^2\cos\theta_w
    \sin\theta_w} (\vec E_\gamma \cdot  \vec H_\z + \vec E_\z \cdot  \vec
  H_\gamma) \:.
\end{equation}
As we will see this leads to the effects similar to those, described in
Section~\ref{sec:observ-inflow}.

\subsection{Static electric field in a capacitor in a magnetic field}
\label{sec:change-E}

Consider again a capacitor in a strong magnetic field $\vec H$, such that
electric field $\vec E$, created by the capacitor, is in the same direction
$x$.  In such a background the current $j^\mu_\z$ is not
conserved~\eqref{eq:96}.  This implies existence of Chern-Simons terms.
These terms creates an inflow $\textsc{Z}$ current, generating anomalous
density of \textsc{Z} charge on the brane. Such a distribution of \textsc{Z}
charge creates \textsc{Z} field and non-trivial $\gamma\z$ background (as in
the right hand side of eq.~(\ref{eq:97})). As a result an inflow of
\emph{electro-magnetic} current appears.  Such an inflow creates anomalous
distribution of electric charge on the brane and modifies electric field
inside and outside the capacitor. Similarly to the case of electrodynamics
(Section~\ref{sec:qed}) we can find this static configuration of an electric
field. It is given by the same expression~(\ref{eq:54}) (see~\cite{anomaly-th}
for details). As a result, distribution of an electric field inside and
outside the capacitor is given by expressions~(\ref{eq:55})--(\ref{eq:57}),
with $m_0$ proportional to the magnetic field $\bH_0$ and anomaly parameter
$\kappa=\kappa_l+3\kappa_q$:
\begin{equation}
  \label{eq:237}
  \boxed{
    m_0^2 = \frac{N_f e^2
      \bH_0}{\pi^2\cos\theta_w\sin\theta_w}\,(\kappa_l+3\kappa_q)
  }\:.
\end{equation}
Notice, that this result does not depend on either mass of the fifth dimension
$\M$ or on the mass of the \textsc{Z}-boson $\m_\z$ (for $\m_\z \ll \M$).
This can be understood as follows: anomaly inflow of the \textsc{Z}-current
creates density of \textsc{Z} charge. The latter, in turn, creates electric
\textsc{Z} field in the 5th direction. This field is, of course, decaying on
the scales larger than $\frac1{\m_z}$.  However the inflow comes from the
region~(\ref{eq:53}) which is much smaller than $\frac1{\m_z}$.  As a result,
in the leading order in $\frac{\m_\z}{\M}$ mass of the \textsc{Z} boson does
not modify the effect.

\section{Experimental detection of anomaly inflow from extra dimensions}
\label{sec:numbers}

Next, we propose an experiment one can conduct to determine the anomaly inflow
and thus the presence of extra dimensions.  As we have already seen in the
previous sections, regardless of the model we are using, there is a common
effect: distribution of an electric field in a capacitor changes, if the
capacitor is placed in a parallel magnetic field. In particular, the electric
field would appear \emph{outside} the plates of the ideal capacitor with the
infinite plates.  A detection of such a field serves as a signature of extra
dimension(s) and provides a direct measurement of the parameter $\kappa_0$. We
should stress once again that this is a \emph{low-energy
  signature of extra dimensions}.

We remind, that in this paper we discuss an (idealized) capacitor: two charged
plates, parallel to each other and separated by the distance $d$ much less
than their characteristic sizes. Such plates create an electric field $\bE_0$
between themselves. The capacitor is placed into a magnetic field $\bH_0$,
parallel to its electric field. As an estimate we will take $\bH_0\sim
10^4$~Gauss (equivalently, 10~Tesla) -- which is of the order of highest
stable magnetic fields, achievable in laboratory (see e.g.~\cite{nhmfl}).

To find out whether the described effect can be observed, we should estimate
the value of $\kappa_0$.  There are several quantities, which can be directly
measured in experiments and whose non-zero values can serve as potential
anomaly mismatch parameters.  They include: the charge of the neutrino, the
charge difference between the electron and the proton, etc.

\begin{table}[t]
\newcounter{line}
\setcounter{line}{1}
  \centering
  \begin{tabular}[c]{|c|c|c|c|c|c|c|}
\hline
Model & $\kappa_0$ & $m_0$, GeV & $\tau_0$, sec & $L_0$, cm & ${\bE_{outside}}/{\bE_0}$ &
$\bE_{outside}$, Volt/m \\
\hline
\hline
\footnotesize new generation & 1 &  $10^{-8}$ & $3\times 10^{-16}$ & $10^{-5}$ & 0 & 0 \\
\hline
\strut\footnotesize charge $\nu$ & $10^{-15}$ & $4\times 10^{-16}$ & $10^{-8}$ & $3\times 10^2$ &
0.5 & $5\times 10^6$ \\
\hline
\strut \footnotesize electric neutrality & $10^{-21}$ & $4\times 10^{-19}$ & $10^{-5}$ & $3\times
10^5$ & $6\times 10^{-4}$ & $6\times 10^3$ \\
\hline
\strut \footnotesize mass $\gamma$ & $10^{-36}$ & $10^{-26}$ & $3\times 10^2$ & $10^{13}$ & $10^{-11}$ & $6\times 10^{-4}$ \\
\hline
  \end{tabular}
  \caption{\small In this Table we gather the  characteristic values of 
      the time $\tau_0=\frac1{m_0}$ over which the static
      state~(\ref{eq:55})--(\ref{eq:56}) is reached.  The static electric field  $\bE_{outside}$ is almost constant for the 
      distances much smaller than $L_0=\frac1{m_0}$ outside the plates of a capacitor. 
      A value of the non-anomalous electric field $\bE_0$ is taken to be
      $10^7$~Volt/m, $\bH_0 \sim 10^5$~Gauss and the distance between the
      plates $d= 10^2$~cm.} 
  \label{tab:exp}
\end{table}

\noindent\textbf{New generation.}
One of the possible ways to make the electroweak theory anomalous, would be to
have new generation which have only leptons or only quarks in them. Such a
generation could be very massive (say, above electroweak scale) and we would
not see its presence in high-energy experiments. In such a theory $\kappa_0
\sim 1$. According to eq.~(\ref{eq:237}) for the value of magnetic field
$\bH_0 \sim 10^4$~Gauss one would get $m_0\sim 10$~eV (see the first row in
Table~\ref{tab:exp}). Taking distance between the plates $d\sim 1$~cm we get,
according to eqs.~(\ref{eq:55})--(\ref{eq:56}) that \emph{the electric field
  inside the capacitor will be exponentially (as $e^{-10^5}$) close to zero}.
Thus, if such a generation existed, the configuration with non-zero $\vec
E\cdot \vec H$ would be forbidden. The characteristic time for reaching such a
final state $\tau\sim \frac1{m_0}$ would be of the order $10^{-16}$~sec. Thus,
we see that existence of such a generation can be considered experimentally
excluded, as its presence would prohibit existence of parallel electric and
magnetic field.  This result is independent on neither energy scale of an
extra dimension, nor the mass of this additional generation.

\smallskip

\noindent\textbf{Charged neutrino.} Another possibility for anomalous theory would be
to add a small electric charge to the left-handed neutrino: $e_\nu =
\kappa_{0,\nu} e$.  This would make the electrodynamics anomalous. We add a
mass term $\lambda\chi\bar\nu^c \nu$, which does not require a charged Higgs
field and thus in such a model photon remains massless.  The modern
restrictions on the charge of the neutrino then give~\cite{pdg}
$\kappa_{0,\nu} < 10^{-15}\:$.
Again, substituting this $\kappa_{0,\nu}$ into eq.~(\ref{eq:49}), we see that
after the time of the order $10^{-8}$~second the system would reach a state,
in which an electric field, as measured outside the capacitor, close to the
plates (closer than $L_0\sim10^2$~cm, see line 2 in the Table~\ref{tab:exp}),
will be of the same order as an electric field inside. Thus, this effect is
very strong and, if present, can easily be detected and serve as a signature
of extra dimensions.

\smallskip

\noindent\textbf{Electric neutrality of matter.} As discussed in
Section~\ref{sec:anom-sm} in SM current restriction on an
anomaly mismatch parameter comes from the measured bound between electric
charges of electron and proton, which gives~\cite{charge-diff}
$\kappa_{0,\textsc{e}\textsc{w}}<10^{-21}\:$.
As one can see from the third row of the Table~\ref{tab:exp}, an electric
field outside the capacitor~(\ref{eq:57}) would be 4 orders of magnitude lower
from its initial value inside. The field inside will be lowered by
approximately the same amount (according to eq.~(\ref{eq:55})). Whether such a
change in electric field can be determined, depends on the actual setup of the
experiment. If we estimate the value of $\bE_0$ by
$\sim10^7$~Volt/meter,\footnote{Electric fields of this order are used in
  linear accelerators, such as the one in Stanford Linear
  Accelerator  Center~\cite{slac}.} %
then an electric field outside the capacitor will be of the order
$10^4$~Volt/meter and can easily be measured. 

\smallskip

\noindent\textbf{Massive electrodynamics.} A modification of electrodynamics in
which left and right moving particles have different charges is another
example of anomalous theory. If left and right moving particles have different
charges, they can acquire mass only via the Higgs mechanism.  Since the mass
term $f\, \bar\psi_L \psi_R\phi + c.c.$ is gauge invariant, the Higgs field
should also be charged with the charge $e_\phi = e_L - e_R$.  The charge of
the Higgs particle is proportional to $\kappa_0$ ($e_\phi \sim e\kappa_0$) and
is very small. The experimental restrictions on such \emph{millicharged
  particles} allow for their existence in the mass range from electron-volts
up to hundreds of GeV (see, e.g.,~\cite{milli}). The presence of the charged
Higgs field in turn implies that the photon acquires mass $m^2_\gamma =
\la\phi\ra^2 (e_L-e_R)^2$.  The most stringent experimental restrictions on
the mass of the photon is $m_\gamma < 3\times 10^{-27}$~eV~\cite{chibisov}.
Taking an estimate for the Higgs's VEV to be of the order 1~GeV (this value
should be much larger than the mass of the fermions), we find the restrictions
for the anomaly parameter to be $\kappa_{0,\gamma} < 10^{-36}\:$.  From the
last row in Table~\ref{tab:exp} we see that the effect would be much smaller
than in previous cases and hard to determine.\footnote{It was argued recently
  in~\cite{dvali} that interpretation of experiments leading to this bound (as
  well as to the less stringent one ($6\times10^{-17}$~eV), cited
  in~\cite{pdg}) may be wrong. The new bound would be $m_\gamma \lesssim
  10^{-14}$~eV. This would provide an estimate for $\kappa_0$ similar to that,
  coming from electric neutrality of the matter.\label{fn:2}}


\smallskip

\noindent\textbf{Restriction on anomaly mismatch coefficient from atomic measurements.}
In the experiment described above we had macroscopic configuration with
non-zero $\vec E\cdot\vec H$, created by a capacitor placed in a solenoid. On
the other hand, any charged particles with spin creates a field configuration
with non-zero $\vec E\cdot \vec H$. To estimate this effect, we start from the
quasi-classical expressions for electric and magnetic fields, created by a
particle with an electric charge $e$ and magnetic moment $\mu=\frac em$
\begin{equation}
  \label{eq:77}
  \vec E = e\frac{\vec r}{r^3};\quad \vec H = \frac{3(\vec \mu\cdot \vec
    r)\,\vec r  - r^2\vec \mu}{r^5}
  \Longrightarrow    \vec E\cdot \vec H= \frac{2(\vec\mu\cdot \vec r)e}{r^6}\:.
\end{equation}
Thus, in the region of the space where expression~(\ref{eq:77}) is non-zero,
inflow current (analogous to that of Section~\ref{sec:observ-inflow}) creates
non-zero anomalous charge density $\rho$. As a result, an electric field
configuration will be significantly modified from the usual Coulomb to Yukawa
form~\cite{anomaly-th} at the distances large than $m_0^{-1}$, with $m_0$
given by an analog of formula~(\ref{eq:49}) for $H\sim e m^2$:
\begin{equation}
  \label{eq:98}
  m_0 \sim m  \sqrt{\kappa_0 e^3} \:.
\end{equation}
This means that the electric charge of the particle \emph{gets completely
  screened} and the total amount of an anomalous charge which appeared on the
brane is equal to the charge of the particle. Such modification of electric
field could influence, for example, atomic measurements. This provides a
restriction on anomaly mismatch coefficient $\kappa_0$: the characteristic
scale $m_0^{-1}$ over which the electric field changes, should be much bigger
than the atomic scale $r_a\sim\frac{1}{e^ 2 m}$. In particular, their ratio:
$r_a m_0=\sqrt{\frac{\kappa_0}{e}}$ should be smaller than the precision with
which the fine structure constant is measured (which is currently
$10^{-10}$~\cite{pdg}). From Table~\ref{tab:exp} one can see that this
restriction allows for $\kappa_0 \lesssim 10^{-21}$, excluding first two rows
in Table~\ref{tab:exp}, but allowing for the two last ones (see also
footnote~\vref{fn:2}).

\section{Initial stage of anomaly inflow}
\label{sec:old-part}

One can ask what happens on the time scales much smaller than $\tau_0$,
specified in the Table~\ref{tab:exp}. This question was analyzed
in~\cite{anomaly-th}. Below we sketch the results and discuss whether they
have any experimental significance.


At initial stage of the inflow one can treat it by perturbation theory in
$\kappa_0$ and consider linear in time response.  In case of the capacitor,
the answer is qualitatively clear: electric charge would appear inside of the
capacitor (in the region where $\vec \bE\cdot \vec \bH \neq 0$ initially) and
changes of electric field will be of the order $\Delta \bE \sim
\frac{t}{\tau_0} \bE_0$. The exact computation in electrodynamics show that
the actual answer for an electric field, \emph{appearing outside the
  capacitor} for times $t\ll \tau_0$ is: $\Delta\bE = \frac{t}{\tau_0 (m_0
  d)}\bE_0$. This means, that if one did not observe any effect over the time
$t_{exp}$, this time provides a lower bound on the characteristic time
$\tau_0$ and hence on the anomaly mismatch coefficient.


As we have already mentioned above, any charged particles with spin creates a
field configuration with non-zero $\vec E\cdot \vec H$~(\ref{eq:77}). The
anomaly inflow in this case leads to an appearance of an anomalous dipole
moment of the particle.  Indeed, in the region of the space where
expression~(\ref{eq:77}) is non-zero, inflow current (analogous to that of
Section~\ref{sec:observ-inflow}) creates non-zero anomalous charge density
$\rho$.  Such density will be positive inside one hemisphere, surrounding a
particle, and negative in another one. As a result, the total charge of the
particle does not change.  However due to the inflow any particle acquires an
anomalous \emph{electric dipole moment}:
  $d_\anom \sim e{(\rho \,r_c^3)}\, {r_c}$.
  Here $r_c$ is a Compton radius of the particle $r_c \sim \frac 1m$ ($m$
  being the mass of a particle).  Anomalous charge density grows linearly in
  time, being $\rho_\anom \sim t E H$, with electric and magnetic fields
  estimated by $E\sim \frac{e}{r_c^2}=e\, m^2$ and $H\sim \frac{e}{m \,r_c^3}
  = e m^2$.  Substituting all these values into expression for $d_\anom$ we
  get:
\begin{equation}
  \label{eq:79}
  d_\anom\sim e^3  \kappa_0\,t\:,
\end{equation}
i.e. as a consequence of anomaly inflow a particle acquires a dipole moment,
which has the absolute value growing with time.  A similar effect exists in
SM. Namely, a background with non-zero $\vec E_\gamma \cdot
\vec H_\z+\vec E_\z \cdot \vec H_\gamma$ can be created around any particle
which has a spin and also electric and $z$-charges. The dipole moment there
can be estimated as
\begin{equation}
  \label{eq:104}
 d_\anom \sim \frac{e^2 \kappa_0}{\cos\theta_w\sin\theta_w}
  \left(\frac{m}{\m_\z}\right)^2\: t \:,
\end{equation}
(where $\theta_w$ is the Weinberg angle).  We see, that unlike the final stage
configuration, this result depends on $\m_\z$.

These results hold only in a perturbation theory, valid for times $t\ll \frac
1{m_{0}}$, where $m_0$ is given by~(\ref{eq:98}).  For times, much bigger that
this characteristic time, field configuration around the particle approaches
static solution. Static equations are symmetric with respect to inversion
$\vec x \to - \vec x$, therefore, there cannot be any dipole moment for $t\gg
\tau_{particle}$.  

Can an anomalous dipole moments of a particle be observed? Linear stage of
anomaly inflow has a characteristic time
\begin{equation}
  \label{eq:238}
  \tau_{particle} \sim \frac1{m \sqrt{\kappa_0 e^3}}
\end{equation}
(where $m$ is the mass of a particle). For a particle to have a dipole moment,
this time should be much greater than the particle observation time.  Taking
as an estimate for the $\tau_{particle}$ for the electron the age of the
Universe ($1.3\times 10^{10}$~years) we get
\begin{equation}
  \label{eq:80}
  \kappa_0  <
  \frac1{m^2\tau_e^2 e^3}\sim 10^{-69} \:.
\end{equation}
Substituting this value into the expression for dipole moment~(\ref{eq:104})
we obtain dipole moment $d \sim 10^{-54}\,e\cdot\mathrm{cm}$ which is many
orders of magnitude below the currently observed experimental value $d_e <
10^{-27}\:e\cdot$cm~\cite{pdg}. Similar computation for the proton would give
$\kappa_0 < 10^{-76}$ and thus even smaller dipole moment (the currently
observed value $d_p < 10^{-23}\,e\cdot\mathrm{cm}$).

As a different example let us consider restrictions on the dipole moment of a
muon. In this case we can take muon life-time $\tau_\mu \sim 2\times
10^{-6}$~sec as an estimate for the characteristic time~(\ref{eq:238}). This
gives $\kappa_0 < 10^{-35}$.  However, the dipole moment, computed
via~(\ref{eq:104}) is of the order $d\sim 10^{-35}\,e\cdot\mathrm{cm}$ which
is still many orders of magnitude smaller than the experimental data for muon
$d_\mu < 3\times 10^{-19}\:e\:$cm.

We could take a different point of view: using eqs.~(\ref{eq:238})
and~(\ref{eq:104}) we can compute $\kappa_0$ necessary to obtain an observable
dipole moment $d_p \sim 10^{-23}\,e\cdot\mathrm{cm}$ for, say, proton. This
gives us $\kappa_0 \sim 10^{-4}$. However, the results of
Section~\ref{sec:numbers} show, that such a $\kappa_0$ would lead to the
complete screening of an electric field in the capacitor in a magnetic field
over the time $\tau_0 \sim 10^{-14}$~sec. Therefore, such possibility can be
considered experimentally excluded.

Thus, we see that anomalous dipole moment arising at the linear stage of
anomaly inflow of the particle cannot be detected regardless of the value of
anomaly mismatch coefficient. Although the value of anomalous dipole moment is
directly proportional to the $\sqrt{\kappa_0}$, the time of the linear stage
is inversely proportional to the $\kappa_0$. Thus, for the values of
$\kappa_0$ providing dipole moments large enough to be measured, duration of
the linear stage of anomaly inflow is too short to conduct an experiment.

\section{Conclusion}
\label{sec:disc}

In this paper we suggested an experiment which can serve as a signature of the
presence of extra dimensions. This experiment is based on the
observation~\cite{anomaly-th} that in an anomalous theory, embedded into a
higher-dimensional one, an electric charge placed in a magnetic field gets
screened with the radius of the screening proportional to the anomaly
coefficient in a theory. We should stress that such an effect \emph{would not
  be present} in a purely 4-dimensional theory.
%
Outcome of proposed experiment does not depend on the scale of extra
dimensions.

Modern experimental data shows that our world is non-anomalous with a very
high precision (experiments include electric neutrality of the matter, absence
of the photon mass, etc.). However, we have suggested a setup in which
the effect can be pronounced enough to be detected even if current experimental restrictions on the
anomaly parameters are satisfied.

The negative result would serve as a restriction on such parameters
as charge of neutrino and electric neutrality of the matter. However, this
statement is true only in the class of models with extra dimensions, similar
to the ones, described here.

Another effect of anomaly inflow is an appearance of a time-dependent dipole
moment of a particle. Although we estimated that such an effect cannot be
observed directly, it would be interesting to see whether its presence can
have a signature in various high-energy processes.
Certainly, there may exist other low-energy (and high-energy) signatures of
anomaly inflow. Some of them may provide stronger restrictions on anomaly
parameters and better experimental setup than the ones, discussed above. We
plan to address these issues elsewhere.

\subsection*{Acknowledgments}

We thank I.~Antoniadis, S.~Dubovsky, J.~Harvey, S.~Khlebnikov for useful
comments.  The work of A.B. and M.S. was supported by the Swiss Science
Foundation. O.R. would like to acknowledge the partial support of the European
Research Training Network  "ForcesUniverse" (contract 005104) and warm
hospitality of Ecole Polytechnique F\'ed\'erale de Lausanne where a part of
this work was done.  A.B. would like to acknowledge the hospitality of
Institut des Hautes \'Etudes Scientifiques.

\end{document}